\documentstyle[12pt]{article}
\catcode`\@=11
\def \be  {\begin{equation}}
\def \ee  {\end{equation}}
\def \beq  {\begin{equation}}
\def \eeq  {\end{equation}}
\def \ba  {\begin{eqnarray}}
\def \ea  {\end{eqnarray}}
\def \baa {\begin{eqnarray*}}
\def \eaa {\end{eqnarray*}}
\def\bea{\begin{eqnarray}}
\def\eea{\end{eqnarray}}

\def\beq{\begin{equation}}
\def\eeq{\end{equation}}
\def\ba{\beq\new\begin{array}{c}}
\def\ea{\end{array}\eeq}
\def\be{\ba}
\def\ee{\ea}

\begin{document}


\setcounter{footnote}0


\begin{center}

\begin{flushright}
{ ITEP/TH-24/04}\\
{ FTPI-04/36}\\
{NSF-KITP-04-126}
\end{flushright}

\vspace{1.3cm}

\begin{center}
{\Large \bf  From effective actions to the background geometry }
\end{center}

\vspace{0.3cm}

\centerline{\large   A. Gorsky ${}^{a,c}$,   V. Lysov ${}^{a,b}$}

\vskip 0.5cm
\centerline{${}^a$ {\em Institute of Theoretical and
Experimental Physics, Moscow 117259, Russia}}
\vskip 0.2cm
\centerline{${}^{b}$ {\em Institute of Physics and Technology,
Moscow, Russia}}
\vskip 0.2cm
\centerline{${}^c$ {\em
W. Fine Theoretical Physics Institute,
University of Minnesota,
Minneapolis, USA}
}
\vskip 2cm

\begin{abstract}

We discuss how the background geometry can be traced from  the
one-loop effective
actions in nonsupersymmetric theories in the external abelian fields. It is shown that
upon the proper identification of the Schwinger parameter
the Heisenberg-Euler
abelian effective action involves  the integration over
the $AdS_3$, $S_3$ and $T^{*}S^3$  geometries,
depending on the type of the external field.
The interpretation of the
effective action in the sefdual field in terms
of  the topological strings is found and the
corresponding matrix model description  is suggested. It is shown that
the low energy abelian
MHV one-loop amplitudes are expressed in terms
of the type B topological string amplitudes in mirror to $T^{*}S^3$
manifold.
We also make some comments on the relation between  the imaginary part of the effective
action and the branes in SU(2) as well as on the geometry of the contours
relevant for the path integral.

\end{abstract}
\end{center}
\section{Introduction}

The duality between  gauge theories and strings is presently under the
attacks from the different directions. One of the most promising
approaches involves   the duality between N=4 SYM
theory and IIB string theory in $AdS_5\times S_5$ geometry with
the additional flux of the four-form field \cite{maldacena}. The string
tension is  proportional to  $\sqrt{\lambda}$ where
$\lambda=g_{YM}^2N$ is t'Hooft coupling hence the strong coupling
limit in Yang-Mills theory can be treated almost classically on
the stringy side. On the other hand the weak coupling limit on the
gauge theory side in principle deserves the knowledge of the full quantum
description of the string in this background - the problem which
has not been solved yet.

If the correspondence is true it is desirable to derive the $AdS$ type
geometries from the first principle starting  from the perturbative
Feynmann diagrams. The interesting step in this direction  has
been made in \cite{gopakumar} where  the first quantized picture
was used to argue that the one-loop two and three point functions in
the scalar theory can be naturally described in terms of the
bulk-to-boundary propagators in $AdS_5$ integrated over position
of the point in the bulk. It turned out that in the first
quantized picture the Schwinger parameter serves as the radial
coordinate in $AdS_5$ hence the integration over this variable
corresponds to the integration over the interior of the fifth dimension.
However starting from the four point function the situation appears
to be more subtle and fully satisfactory picture is absent.

Therefore it is natural to clarify the situation using some
different tractable object. The good candidate to work with is the
effective action in the external field. It effectively involves
the arbitrary order in the coupling constant and on the other hand the
explicit answer can be derived. We shall consider such effective
action from the background geometry perspective and
shall argue that peculiar geometries emerge
in a very natural way. In the particular case of the abelian
theory with the constant electric and magnetic fields effective
action will be formulated in terms of the $AdS_3$ and  $S^3$
geometries where one of the  coordinates shall be identified with
the Schwinger parameter similar to discussion in \cite{gopakumar}.
Note that the first quantized picture we shall use below has been
applied for the effective actions long
time ago \cite{strassler,schubert}.

More deep motivation for the choice of the effective action in the
external field as the convenient object to capture the geometry is as follows. The
experience of the work with the supersymmetric theories suggests
that a kind of the refined "holomorphic" object encodes the information
about the external geometry. In the  N=2 theories this role
is played by the effective prepotential while in the  N=1 theory the effective
superpotential is the relevant object. In both cases the prepotential
and superpotential can be read off from some Riemann surface
embedded into the  three dimensional noncompact Calabi-Yau manifold.
The most efficient computational tool to get the explicit
form of the effective actions in SUSY theories involves the
topological strings (see \cite{marino,nv} for review). The
open string topological A picture involves Chern-Simons type
description while type B topological open string provides
the corresponding matrix model. Both pictures
get mapped into the closed string geometry upon the large N transition.

In this paper we shall consider nonsupersymmetric theories so at the
first glance there is no such refined object at all and the situation looks
hopeless. However it turns out that at least some part of the
geometry can be read off from the effective actions in this case too.
The proper geometric interpretation of the Schwinger parameter
provides the important starting point. In the case of the selfdual
background we will be able to go further and apply similar
topological string ideas in the nonsupersymmetric case. The reason
for such possibility can be naturally attributed to the residual
supersymmetry known in the selfdual background. In fact we shall try
to combine the geometrical interpretation of the Schwinger
parameter with the picture familiar from the topological strings.

In the selfdual case we shall use the relation between the large N
Chern-Simons theory and the Schwinger type calculations
discussed in the context of the topological strings \cite{gv9802}.
The gauge theory in the type A picture is realized on the worldvolume
of D6 branes wrapped around $S^3$ considered as the Lagrangian
submanifold in $T^{*}S^3$. Chern-Simons theory lives on $S^3$
and upon the  large N transition it is dual to the theory on the resolved conifold
with fluxes instead of branes \cite{gv9811}. The logarithm of the CS partition function
at large N turns out to coincide with the scalar QED effective action
in the selfdual field. This relation appears to be consistent with  the
identification of the Schwinger parameter in the wrapped picture.
The similar representation will be found for the  pure electric and magnetic
backgrounds.

Surprisingly enough we can formulate the matrix model picture
for the effective action in the selfdual background at least at one loop. It is based on the
matrix model representation of Chern-Simons theory on $S^3$ found
in \cite{marino1,marino2} which corresponds to the
geometry of the type B model identified with the mirror to $T^{*}S^3$.
It turns out that the potential in the matrix model
at least in one formulation involves the double trace terms.
Let us emphasize that
this matrix model is the nonsupersymmetric counterpart of the
matrix model found by Dijkgraaf and Vafa \cite{diva} which describes the effective
superpotential in N=1 SUSY YM theory. We shall make some comments concerning the
proper interpretation of this matrix model in terms of the spectrum
of the Dirac operator in the background fields.

In principle one could also consider a kind of the probe  picture on the brane
and use the representation of the  effective action
as the sum with the proper weights over the Wilson loops with
the different boundary contours. These contours  can be considered as
the boundaries of the string worldsheets embedded into the ambient geometry.
Roughly speaking in the probe picture the Schwinger parameter
measures the lengths of the
boundary contours
and the effective action from the probe brane
perspective can be considered as the back reaction of the string
ending on it.
To recognize the background geometry for the probe  brane we first
identify the integrands in the effective actions as the
propagators of the particles with the different masses in $AdS_3$ and
$S_3$ metrics. Therefore  these geometries are expected to serve  as
the background  for the probe D3 brane with the constant fields indeed.
Such geometries are natural to emerge since, for instance, magnetic
field amounts from the density of the D1 strings on the D3 worldvolume
which yields the $AdS_3$ geometry around.

It is known from the textbooks that  the one loop abelian effective action serves as the
generating function for the low energy one-loop photon amplitudes.
In particular the effective action in the selfdual background yields
one loop MHV amplitudes with all "+" or "-" photons \cite{shubert2}. As a byproduct
of our analysis we shall be able to develop an interesting  type B topological string
representation of such MHV amplitudes very much in the spirit
of the recent discussion of nonabelian gluonic MHV amplitudes \cite{wittentw, one-loop}.
In our case the mirror of $T^{*}S^3$ plays the role of the "twistor"
manifold. Moreover we shall derive the topological string interpretation
for the generating function for MHV amplitudes with arbitrary number
of photons involved. It turns out that such MHV amplitudes can be also
derived from the perturbative expansions in Chern-Simons theory on $S^3$
or in the corresponding matrix model.

The effective action in the electric field develops the imaginary part corresponding
to the nonperturbative pair creation. We shall briefly discuss these
issues from the point of view of our realization
of the Schwinger parameter and make some relation with the quantization
of the brane radii on SU(2) group manifoilds. We shall also present
qualitative arguments implying
that the curves with cusps are important in the
path integral.

The paper is organized as follows. In Section 2 we shall briefly
review Gopakumar's arguments concerning the identification of the
Schwinger parameters with the radial coordinate in $AdS_5$. In
Section 3 we will show that the propagators in $AdS_3$ and  $S^3$
type geometry are involved in the effective actions in the
external abelian field. In Section 4 we shall find the precise
background geometry for the selfdual external field and discuss
the  matrix models which
substitute Dijkgraaf-Vafa model for nonsupersymmetric case
at one loop level.
In Section 5 we discuss the topological string
representation for the abelian low energy one-loop MHV amplitudes.
The relation between the spherical branes in $S^3$ and
the pair production in the electric field is discused in Section
6. Section 7 concerns the role of
the contours with cusps in the path integral representation of the
effective action. Some open questions and possible generalizations can be
found in the last Section.

\section{Free fields and AdS}

In this section we briefly review the arguments of Gopakumar
\cite{gopakumar} concerning the representation of the loop
diagrams in the scalar theory in four dimensions in terms of the
tree diagrams  in $AdS_5$. He adopted the first quantized language
for the calculation of two and three point functions  and, for
instance, the two point function can be presented as
\beq
\Gamma(k)= \int _{0}^{\infty} \frac{d\tau \tau^2}{\tau^{d/2+1}}
\int d\alpha e^{-\tau \alpha(1-\alpha)k^2}
\eeq
where the
exponential factor comes from the worldline correlator of two
vertex operators
\beq
<e^{ikx(\tau _{1})}e^{-ikx(\tau
_{2})}>=e^{-k^2G(\tau_1,\tau_2)}
\eeq
and the worldline propagators
reads as
\beq
G(\tau_1,\tau_2)=-\frac{\tau_{12}(\tau-\tau_{12})}{\tau}
\eeq
The
variable $\tau$ corresponds to the invariant length of the path
$\tau=\int e(t)dt$ in the Polyakov like formulation of the
particle action
\beq
S=\int_{0}^{1}(e^{-1}\dot{x}^2 +m^2e) dt
\eeq
It is the particle counterpart of the stringy Liouville mode and
therefore similar to the string case is a good candidate for the
additional fifth dimension.

In the space representation  the answer has the structure of
the heat kernel
\beq
<x|e^{t\Delta}|y>=\frac{1}{(4\pi
t)^{d/2}}e^{-\frac{(x-y)^2}{4t}}
\eeq
and therefore using the
relation between the heat kernel and the bulk-boundary propagator
\beq
K(t)=\int d\rho \rho^{d/2 -3}e^{-\rho}e^{\frac{t\Delta}{4\rho}}
\eeq
in can be brought modulo the overall factor to the form
\beq
\Gamma(x_1,x_2)= \int dz_0 z_0^{-(d+1)}K(x_1,z)K(x_2,z)
\eeq
corresponding to the tree representation of the two-point function in
$AdS_5$. Let us note that in this two-point case we can just
consider the product of two free propagators from the very
beginning
\beq
G(x_1-x_2)= \int ds
s^{-(d/2+1)}e^{-\frac{(x_1-x_2)^2}{4s}}
\eeq
and introduce the new
parameter
\beq
s_{tot}^{-1}=s_1^{-1} + s_2^{-1}
\eeq
which allow
to treat the product in terms of $AdS_5$ geometry.

These simple arguments were shown to work for the three point
functions  but some modification for the higher point
functions is needed. It is worth noting that the effective radius
in $AdS_5$ geometry in this case is not related to the gauge
coupling constant contrary to the standard gauge/string
correspondence where $\sqrt{g^2 N_c}=\frac{R_{AdS}}{\alpha '}$.

\section{Effective actions in the external fields and the propagators in the curved
background}
\subsection{Propagators in $AdS_3$ and $S^3$}

After the reviewing the free field case let us turn now to our
main example - the theory in the external electromagnetic field. We
shall try to recognize the  background geometry in the loop
calculations in the external field. Let us note that the external
field to some extend provides the IR regularization of the theory.
On the other hand as we shall see later it gives rise to the effective
$AdS_3 $ or $S^3$  type geometry with the radii related to the coupling
constant. Moreover it shall be clear that these nontrivial geometries
emerge after the summation over all orders in the external field
and can not be seen at any fixed order.

Let us start with the effective action for  the fermion with the mass m arising from
the determinant in the external field
\beq
S_{eff}=logdet(i\partial -eA -m)=\int\frac{dT}{T}exp(-Tm)\int
dxTr<x|exp(-tD)|x>
\eeq
The equivalent representation in the first
quantized picture looks as follows
\beq
S_{eff}=\sum_{paths
C}exp(-mL(C))exp(-i\Phi(C))<W(C)>
\eeq
where
\beq
W(C)=TrPexp(ie\oint_{C} Adx)
 \eeq
 and the factor $exp(-i\Phi(C))$
is responsible for the spin impact on the path integral. For
instance in two dimensions it just counts the number of
selfintersections $\nu(C)$ of the contour C with the weight
$(-1)^{\nu(c)}$. In higher dimensions it can be described via a
Wess-Zumino terms on  $CP_n$ geometry, for instance
$CP_1$ for the fermions in three  dimensions \cite{polyakov} and
$CP_3$ for the four dimensional fermions \cite{kor}. This form
involves the summation over the Wilson loops and is appropriate
for the derivation of the stringy picture. In particular, in the
strong coupling limit each Wilson loop can be considered as the
boundary of the open string worldvolume extended into the radial $AdS$
direction.

The general Euler-Heisenberg action for the fermions when both
magnetic and electric fields are involved reads as (see, for instance, the recent
review \cite{dunne})
\beq
\label{both}
L_{eff}=\int_{0}^{\infty}
\frac{ds}{s^3}e^{-sm^2} [esa ctg(esa)\times esb cth(esb)
-1-\frac{e^2s^2}{3}(b^2-a^2)]
\eeq
where $a,b$ are the standard invariants $ab=E\cdot H$,
$a^2-b^2=E^2 -H^2$. In what follows we shall   omit the
subtraction terms which can be restored in a trivial manner.
The similar expressions are known for the
particles of the different spins, for instance, the explicit
calculation amounts to the following effective action for the spin
S particle in the electric field
\beq
\label{arbit}
L_{eff,scal}=\int_{0}^{\infty} \frac{ds}{s^3}
\frac{eEs}{sin(eEs)}\frac{sin(2S+1)eEs}{sin(eEs)} e^{-sm^2}
\eeq
It reduces for the  fermion to
\beq
L_{eff,f}=\int_{0}^{\infty}
\frac{ds}{s^3}seEctg(eEs)e^{-sm^2}
\eeq
and  for the scalar to
\beq
L_{eff,scal}=\int_{0}^{\infty} \frac{ds}{s^3}
\frac{eEs}{sin(eEs)}e^{-sm^2}
\eeq

It is also convenient to
present  the expression for the effective action
for spin 1/2 particle in the constant selfdual field G
of electric type which we shall
intensively use in what follows
\beq
L_{eff}=\int_{0}^{\infty}
\frac{ds}{s^3}exp^{-sm^2} (esGcth(esG))^2
\eeq
while for the scalar we have
\beq
L_{eff}=\int_{0}^{\infty}
\frac{ds}{s^3}exp^{-sm^2} (\frac{esG}{sin(esG)})^2
\eeq
Let us emphasize that there are two types of the selfdual
background;  one is similar to the electric case while the
second to the magnetic one. There is the standard pair production
mechanism in the electric version of the selfdual field.

Let us argue now that the effective action naturally involves
the propagators in $AdS_3$ and $S_3$ geometry very much in a spirit of
Gopakumar's calculation. Consider the propagator of the
massless particle on the SL(2,R) group manifold which coincides with $AdS_3$.
The propagator is defined as
\beq
G(x,y)=-i\int d\tau <x|e^{-\tau
\Delta_{ads_3}}|y>
\eeq
where $\Delta_{ads_3}$ in the
Laplace-Beltrami operator on SL(2,R). The propagator can be
calculated using the expansion of the
transition amplitude over the characters of SL(2,R)
unitary continuous  representations. There are
fundamental series of such representations with $j=-1/2 +i\nu/2$
and complementary one with $-1<j<0$.
Summing over two series one arrives at the following expression
\beq
G(x,y)= \theta coth \theta
\eeq
were $\theta$ is
defined via
\beq
cosh \theta= \frac{1}{2}tr( g_x^{-1}g_y)
\eeq
and
the following parametrization of the group manifold is assumed
\bea g_x= \frac{1}{z_0} \left(
\begin{array}{cc}
  z_1-z_2 & 1 \\
  z_1^2-z_2^2-z_0^2 & z_1+z_2 \\
\end{array}
\right)
 \eea
In this parametrization we have the metric in the standard form
\beq
 ds^2 =tr( g_x^{-1}dg_x)^2=\frac{dz_0^2-dz_1^2+dz_2^2}{z_0^2}
\eeq
Hence we immediately recognize the ingredient of the effective
action  for the fermion in the external magnetic field as the
propagator of the massless mode in $AdS_3$ background. To compare
this observation with variables in \cite{gopakumar} we can identify
\beq
cosh\theta=\frac{1}{2}tr(g_x^{-1}g_y)
=\frac{z_0^2+w_0^2+|z-w|^2}{2z_0w_0}= cosh\xi \eeq For the
massless scalar particle in $d=2$ we obtain $\triangle=0$ and
its propagator on $AdS_3$
\beq
G_{scal}(z_0,w_0,z,w)=(\frac{\xi^2}{\xi^2-1})^\frac{1}{2}=coth\theta
\eeq

The natural question is if the relation to $AdS_3$ we found is
seen for the arbitrary spins as well. Let us consider the exact expression for
the propagator on the AdS introducing the new variable $\mu^2=1-m^2$
 \beq G(\theta)=
\sum_{l=0}^{l=\infty} (\frac{1}{l+1+\mu}+\frac{1}{l+1-\mu})
(e^{(l+1)\theta}-e^{-(l+1)\theta})
\eeq
Removing the
divergency in this sum we obtain the following answer
\beq
G(\theta)=\frac{\theta cosh(\mu\theta)}{sinh\theta}+
\sum_{l=1}^{\mu-1} \frac{sinh (l\theta)}{(\mu-l)sinh \theta}
 \eeq
The leading $\theta \rightarrow \infty $ term in this expression
is close to the leading term in the one-loop external field
expression for the particle of spin S related to the mass as
\beq
(2S)^2= 1-m^2
 \eeq
 which means that the higher spins in the
theory in some sense correspond to the tachyonic modes in the bulk.

The very similar argumentation for the electric field involves the
propagator of the corresponding modes in $S^3$ geometry coinciding
with the SU(2) group manifold. For instance, the propagator of the
massless mode in $S^3$ can be derived  from the summation over the
characters of the unitary irreducible SU(2) representations
\beq
G(\theta)=\sum_{j} \frac{2j+1}{j(j+1)}\chi_{j}(\theta)
\eeq
amounting to
\beq
G(\theta)=\theta cot\theta
\eeq
It is
involved into the effective action for the fermion in the external
electric field where the following parametrization is implied
\beq
cos\theta=\frac{1}{2}trg_x^{-1}g_y
\eeq
The inspection of the
generic effective action  in (E,H) fields shows that the product
of $AdS_3$ and $S^3$ propagators is involved while  for the
external selfdual fields the product of two $AdS_3$ or $S^3$ propagators is
relevant. Hence generically the constant abelian field background
feels the $AdS_3$ or  $S^3$ type gravitational background
around. Similar to Gopakumar's calculation the effective actions involve
the integration over the boundary conditions for propagators in
the background geometry.

Let us briefly discuss the two-point
function in the external field and  consider the simplest
case of the massless scalar theory in the selfdual external field.
This choice is motivated by the very simple form of the scalar
particle propagator in the external field
$F_{\mu\nu}F_{\nu\rho}=-f^2\delta_{\mu \rho}$ in the gauge $xA=0$
\beq
G(x,f)= (\frac{ef}{4\pi})^2 \int_{0}^{\infty}
\frac{d\tau}{sinh^2(ef\tau)} e^{-ef/4 x^2coth(ef\tau)}
\eeq
Hence
the two-point function in $x$ representation is just
\beq
\Pi(x)=\frac{e^{-ef/2 x^2}}{4 \pi^2 x^4}
\eeq

In the free case the bulk-boundary propagators in $AdS_5$
are involved into the two-point functions while in the case of the
external selfdual field the bulk-bulk propagators in $AdS_3$ are
relevant hence it is desirable to explain how $AdS_5$ geometry is
restored if the external field is switched off. To this aim let us
look at the propagator in the external field which after the
change of variables $\xi=ef \cdot coth(ef\tau) $ can be brought to
the following form
\beq
G(x,f)=\int _{ef}^{\infty} d \xi e^{-x^2 \xi}
\eeq
which exactly coincides with the free particle case modulo
the restriction in the integration region. In the external field the
proper time variable $\tau$ corresponding to the radial coordinate
in $AdS$ varies in the limit $\infty>\tau >ef$ instead of $0<\tau
<\infty$ in the free case. Hence the external field plays the role
of the natural regularization with respect to the radial coordinate in
$AdS$ and it is clear that the full $AdS_5$ can not emerge
in this case.

The explicit expression for the two point function for the scalar
field in the external selfdual field in the first quantized picture
\beq
\Pi_2(p)= \int
\frac{dT}{T} \int Dx^\mu e^{\int \dot{x}^2 + e A_\mu\dot{x}^\mu}
\eeq
can be obtained after the simple calculation
\beq \Pi_2(p) =
\int^\infty_0 \frac{(ef)^2dT}{sinh (efT)} \int  \ \ d\alpha  \ \ \
exp\{(\frac{sinh (efT\alpha)}{2ef}-\frac{2sinh^2 (\alpha e f
T/2)}{(ef)^2T})p^2\} \eeq

\subsection{Relation to the 2D Yang-Mills theory}

We have identified the ingredients of the effective action as
the propagators in $AdS_3$ and $S_3$.  Let us represent the
massless propagators in one more way through the two dimensional
Yang-Mills theory on the disc.
Let us consider now  the effective action of the
fermion in the pure electric field. To derive the 2D interpretation of
the effective action let us start with the
representation of the $S^3$ propagator discussed in \cite{bgk}
in terms of  the two-dimensional Yang-Mills theory with
SU(2) gauge group on the  disc.

To make the correspondence exact, one introduces the amplitude of
the two-dimensional Yang-Mills theory on a disk  with radial
coordinate $x^0$, $0 \leq x^0 \leq T$, and angular $x^1$, $0 \leq
x^1 \leq L$, of area ${\cal A} = LT/2$, and a holonomy at its
boundary $C = \partial \Sigma$,
\begin{eqnarray*}
U = P \exp {i \oint_C  A} \, .
\end{eqnarray*}
The partition function on the disk is \cite{migdal}
\begin{equation}
\label{DiskAmplitude} {\cal Z} [U; g^2 {\cal A}] = \int DA_\mu \,
\delta \left( P e^{i \oint_C dx \cdot A (x)},U \right) e^{ -
\frac{1}{g^2} \int_{\Sigma} d^2 x \sqrt {det g_{\mu \nu}} Tr F^2 }
= \sum_{j} (2j + 1) \chi_{j} [U] e^{ - {g^2 {\cal A}} \, j(j +
1)/2 } \, ,
\end{equation}
where and $\chi_j [U]$ are characters for the spin-$j$
representation of the gauge group. Thus,
\beq
{\cal Z} [U; g^2
{\cal A} = 2 \tau] = <-v'| e^{- \tau H} |v> \, , \label{Z-1}
\eeq
where $U = g_{-v'}^{-1} g_v^{\phantom{1}}$ and H is SU(2)
Casimir operator. The propagator of the massless particle on the
SU(2) group manifold is given by the integral of the wave functional in
two-dimensional Yang-Mills theory on the disc with respect to its
area
\beq
G_{S_3} (\theta) =\int_0^\infty d \tau  {\cal Z}
[U; 2 \tau]  \, . \label{cusp=string}
\eeq
Therefore we can represent the effective action
of the fermion in the electric field as the
partition function of the SU(2) YM theory on the disc integrated
over the area of the disc and boundary holonomy
\beq
S_{eff}(E)=\int dsdA
s^{-3}e^{-sm^2}Z_{2dYM}(arccos(esE),A)
\eeq

One could have in mind also the instantonic realization of the
same partition function of the  2d YM partition function
related to its  saturation  by a sum over the
classical saddle points in the path integral
\cite{GroMat94}. The instantons under consideration are solutions
to the Yang-Mills equations of motion on the two-dimensional disc
with the boundary conditions set by the holonomy $tr \, U [A (x^0
= T, x^1)] = 2 \cos \tilde\theta$, where $\tilde\theta =\pi- \theta $.
The classical configurations in
the $A^0 (x^0, x^1)= 0$ gauge correspond to the straight paths
connecting the initial and final points and read in the
topological charge-$\ell$ sector
\beq
A^1_\ell (x^0, x^1) = x^0
(\sigma_3/2) (\tilde\theta + 2 \pi \ell)/{\cal A}.
\eeq
The action
evaluated on these instanton solutions reads
\begin{eqnarray*}
S [A_\ell] = 2 ( \tilde\theta +  2 \pi \ell )^2/(g^2 {\cal A}) \,
,
\end{eqnarray*}
and summation over $l$ amounts to the equivalent representation
for the transition element in YM theory
\beq
<0|e^{i\tau L^2}|v>= \frac{(-i\pi\tau)^{-3/2}}{\pi sin\theta}
\frac{\partial}{\partial\tilde\theta}\sum _{l}
e^{-\frac{i(\tilde\theta +2l\pi)^2}{\tau} -\frac{i\tau}{4}}
\eeq
The representations via characters  and via instantons are related
to each other by the Poisson resummation formula and amount to the same expression
for the partition function.

The picture emerging for the pure magnetic background is very
similar to the electric case with SL(2,R) two dimensional Yang-Mills
theory instead of SU(2) one. The effective action reads now as
\beq
\label{electric} S_{eff}(H)=\int dsdA
s^{-3}e^{-sm^2} Z_{2dYM,SL(2,R)}(A,arccosh(esE))
\eeq
in terms of partition function of Yang-Mills theory. We shall
discuss the emergence and role of the  $AdS_2$ geometry in the magnetic
case later.

In the selfdual electric background case  we have the following
representation for the fermion effective action
\beq
\label{electric} S_{eff}(E)=\int dsdA
s^{-3}e^{-sm^2} Z_{2dYM,SU(2)}^2(A,arccos(esE))
\eeq
In the next section we shall demonstrate that the background
geometry can be derived for the selfdual case very explicitly.

\section{Effective actions in the selfdual field and the topological strings}

\subsection{Scalar case}

In this section we shall find the relation
between  the effective actions in the selfdual background  and
the topological strings very much in the spirit
of the previus discussions in the SUSY case.
We shall exploit
the relation between the large N CS theory on $S^3$
which represents the topological open string A model
and topological string models related to it by
the large N transition or mirror symmetry. Upon the large
N transition the topological A model corresponding to A branes
wrapped around $S^3$ Lagrangian submanifold in $T^{*}S^3$
gets mapped into the topological closed string in the resolved
conifold with fluxes in $P^1$'s instead of branes \cite{gv9811}.
In the mirror dual topological B model  topological branes
are wrapped around 2-cycles in the mirror Calabi-Yau manifold
and upon the large N transition are replaced by the blown up 3-cycles
on the modified Calabi-Yau manifold. The latter has been used in \cite{diva}
to derive the effective superpotential in N=1 SYM theory.

Let us recall the relevant facts concerning CS theory on $S^3$.
The partition function of CS theory on $S^3$ has been calculated
long time ago \cite{periwal} and reads as
\beq
Z_{CS}(N,k,S^3)= e^{\frac{i\pi N(N-1)}{8}} \frac{1}{(N+k)^{N/2}}
{\sqrt \frac{N+k}{N}} \prod_{j=1}^{N-1}(2sin\frac{j\pi}{N+k})^{N-j}
\eeq
The theory can be treated as the open topological
type A string field theory \cite{witten} on $T^{*}S^3$
where the coupling constant is identified with
\beq
g_s=\frac{i}{k+N}.
\eeq
In the brane setup it can be considered as the worldvolume theory on the N topological
A branes wrapped
around the Lagrangian submanifold in $T^*S^3$. The theory implies the
topological expansion
\beq
Z_{CS}(N,k,S^3)=\sum_{g,h}C_{g,h}N^{2-2g}\lambda^{2g-2+h}
\eeq
over the worldvolumes with h holes and g handles, where $\lambda=g_s N$
is the standard t'Hooft coupling. In what follows we shall be
mainly interested in the planar limit
\beq
F_{0}(\lambda)= \sum_{h} F_{0,h}\lambda^h
\eeq

The relation between the  CS partition function and the  topological string amplitudes
has been found in \cite{gv9811} where it was shown that
amplitudes  coincide with the
proper expansion terms of $logZ_{CS}(N,k,S^3)$. On the other hand
such terms can be found via a kind of the Schwinger like loop calculations \cite{gv9802,gv9811}.
From the  dual gravity perspective such terms counts the contribution
to the $W^gR^2$ terms in the N=2 effective action where $W$ corresponds to the
graviphoton background.

We shall use the observation \cite{gv9802} that  the effective action
of the scalar in the abelian selfdual field of electric type
coincides with the $logZ_{CS}(N,k,S^3)$ at $N\rightarrow \infty$
\beq
logZ_{CS}(N=\infty,k,S^3)= \int_{0}^{\infty} dse^{-sk}(\frac{s/2}{sin s/2})^2
\eeq
The
identification of the parameters looks as follows
\beq
k_{CS}= m^2/2ef
\eeq
hence the large level k corresponds to the weak field limit. The limit
$N\rightarrow \infty$ can be thought of as arising from the infinite
number of topological  branes wrapped around $S^3$.  These correspond
to the strings
inside D3 brane in the probe picture which provide the constant selfdual background
field.
The level k expansion of the free energy of CS theory in this limit also coincides with the
genus expansion of the free energy of  c=1
noncritical bosonic string  at the selfdual radius \cite{gv}.
In the framework of type A model open topological string on $T^{*}S^3$
can undergo the topological large N transition to the resolved conifold
without branes. The Kahler class of the corresponding $S^2$ is fixed
by the t'Hooft coupling.

It is most convenient for our purposes to use the mirror type B
topological model which is related to CS theory as follows. First,
let us recall the construction of the manifold mirror to $T^{*}S^3$
which can be presented as
\beq
xu+yv=\mu
\eeq
and can be considered as $T^2\times R$ fibration over $R^3$.
The mirror transforms $T^{*}S^3$ to the blowup of
\beq
xy=(e^u -1)(e^v -1)
\eeq
along the locus $x=y=(e^u -1)=(e^v -1)=0$ by inserting a $P^1$.
The imaginary parts of u and v in the mirror manifold are T dual to
the one-cycles on the torus fiber in A model. After the mirror transform N type B branes
get wrapped around $P^1$.

It was observed in \cite{marino1,marino2} that the large N CS theory
on $S^3$ can be described by the  matrix model defined on the
type B branes wrapped around $P^1$. The theory on their worldvolumes
reads as
\beq
S=\int_{P^1} Trv\bar{D} u
\eeq
where u and v are normal coordinates to the brane worldvolumes.
The theory on $P^1$ should be considered as emerged after the gluing of
two halfs of the sphere with nontrivial map between boundaries.
The gluing operator U fixes the superpotential of the corresponding
matrix model and in the case under consideration it reads as
\beq
U=exp(\frac{1}{g_s}\int_{P^1}\omega Tru^2)
\eeq
where $\omega$ is a (1,1) form. The measure in the corresponding
matrix model turns out to be unitary
\beq
d_{H}u=\prod_{i}du_i (\prod_{i\le j}2sin(\frac{u_i-u_j}{2})^2
\eeq

The matrix model can be solved in the usual way introducing
the density of eigenvalues $\rho(e^u)$ and the corresponding
resolvent $v(e^u)$. In the single cut solution resolvent
fixes the Riemann surface where u and v are defined \cite{marino2}
\beq
(e^v -1)(e^{v+u} -1) +e^t -1=0
\eeq
where t'Hooft parameter is
\beq
t=\frac{1}{2\pi i}\int_{A}vdu
\eeq
This Riemann surface corresponds to the nontrivial part
of the manifold obtained from B model upon the large N transition.

What the coincidence of the effective action in the selfdual
background and $logZ_{CS}(N,k,S^3)$ could teach us? The first lesson
is that now we can  precisely determine the background
geometry from the one-loop effective action
in nonsupersymmetric theory that was one
of our purposes. The case of the scalar effective action
is the most transparent one however later we shall consider
the similar picture for the spinor particle.
The Kahler moduli of $P^1$ in Calabi-Yau can be identified with
the dimensionless combination of the particle mass and the external field
hence the topological amplitudes in type B theory correspond
to the expansion of the effective action in terms of the external
field. The role of the Schwinger parameter is quite clear -
it defines the radii of $S^2$ spheres inside $S^3$ in the type A geometry
while in the Type B model it corresponds to the eigenvalues of the holonomies.

Moreover now we can
develop the matrix model representation for the effective action
of the scalar in the selfdual field in  nonsupersymmetric theory which reads as
\beq
Z_{MM}(g_s)=\frac{1}{VolU(N)}\int d_{H}M exp \frac{1}{2g_{s}}TrM^2
\eeq
where the integral is over the Hermitian matrixes with the Haar measure.
This matrix integral can be traded for the matrix integral with the
standard Hermitian measure but with the additional potential term
corresponding to the double trace operators \cite{marino2}
\beq
Z_{MM}(g_s)=\frac{1}{VolU(N)}\int d_{Herm}M exp(\frac{1}{2g_{s}}TrM^2 +V(M))
\eeq
where
\beq
2V(M)=\sum_{k=1}^{\infty} a_k\sum_{s=0}^{2k}(-1)^s C^{2k}_{s}TrM^s TrM^{2k-s}
\eeq
where $a_k$ can be expressed in terms of Bernoulli numbers $a_k=\frac{B_{2k}}{k(2k)!}$.
The perturbation theory in terms of the vacuum expectation values of the double trace
operators calculated with the Gaussian measure can be developed similar to \cite{marino2}.

Let us emphasize that the very precise background geometry in the selfdual case can be
naturally attributed to the residual supersymmetry of the quantum
fluctuations in the selfdual case. It is natural to expect
that the remarkable relations between two loop and one loop effective actions
found in \cite{dunne2} can be related to the topological string interpretation found in our paper.
This issue is under investigation now.

\subsection{On the matrix model picture}

Since we relate one loop effective action to the matrix integral the standard questions
enhereted from the matrix model technique can be reformulated in  our case.
It is instructive to compare the effective action in the selfdual background  with  the matrix
model description of N=1 theory \cite{diva} and chiral effective action
in QCD.  We could expect two different matrix realizations;  one is the large N
matrix model while the second is Kontsevich type model with finite size matrixes.
We shall indeed see two possible  pictures in what follows and   discuss some key
ingredients like resolvent and loop equation on the matrix model side
from the gauge theory perspective postponing more detailed analysis for the future work.

Let us first remind some points concerning the matrix model description of N=1 theory \cite{diva}.
The corresponding Hermitian matrix model involves the tree superpotential
\beq
Z=\int dM exp\frac{1}{g_s}TrW_{tree}(M)
\eeq
that is the potential of the matrix model coincides with the tree superpotential
and the matrix $M$ is the image of the adjoint chiral field  \cite{diva}.
The effective superpotential can be expressed in terms of the partition function
of this matrix model. The standard matrix model resolvent has the field
theory counterpart
\beq
R(z)=Tr \frac{W^2}{z-\Phi} \leftrightarrow g_sTr \frac{1}{z-M}
\eeq
where $\Phi$ is the adjoint chiral superfield.
The Virasoro constraints get mapped into the Konishi anomalies and
their generalizations \cite{gorsky2,douglas} on the gauge theory side moreover
the loop equation has been
mapped into the chiral ring relations in N=1 SYM theory with the adjoint matter \cite{douglas}.
The loop equation collecting all Virasoro constraints into a single
equation yields the background geometry for the type B topological closed string.

In our nonsupersymmetric case we have no tree superpotential as well as the adjoint
field at all so at the first sight it is unclear what substitutes the resolvent and
anomaly relations in this case properly.  We suggest that  the counterpart of the matrix M
is just the Dirac operator $D(A)$ in the fermion case and $D^2(A)$ gets mapped
into the $M^2$ in the scalar case.
Hence the proper  resolvent on the gauge theory side is
\beq
\frac{\delta S_{eff}}{\delta m}|_{m=z}=G_{F}(z)=Tr\frac{1}{D(A)- z}
\eeq
for fermion  and
\beq
G_{S}(z)=Tr\frac{1}{D^2  - z^2}
\eeq
for scalars and trace is taken over the corresponding Hilbert space.
To get the proper normalization on the matrix model side  we could   use the
Casher-Banks relation for the chiral condensate in the selfdual field
\beq
<\bar{\Psi}{\Psi}>\propto \rho(0)\propto \frac{f^2}{m}
\eeq
and take into account that
\beq
G_{F}(z+i\epsilon) -G_{F}(z-i\epsilon)=\pi\rho(z)
\eeq
Hence it is natural to multiply the matrix model resolvent
by the factor $g_s^2$.

Let us turn to the second ingredient of the matrix model namely the loop
equations which have the chiral ring relations as a gauge theory
counterpart in N=1 case.  The loop equations can be derived
from the change of variables in the matrix integral
\beq
\delta M\propto f(M)
\eeq
and it is convenient to take $f(M)=(z-M)^{-1}$.  On the gauge theory side
the corresponding variations involve the variation of the eigenvalues
of the Dirac operator induced by the variation of the external field.
However more convenient interpretation follows from the
variation of the variable dual to the eigenvalue of the Dirac operator
which is just the Schwinger parameter. The change of variables  in the matrix integral
 can be related to the change of variables $s\rightarrow g(s)$ in the
 path integral.
Therefore the Ward identities in the matrix model
can be translated in the dual representation to the invariance of the integral over the Schwinger parameter
under the change of variable $s\rightarrow g(s)$.
Remind that our starting point
was the identification  of the Schwinger parameter as the coordinate in the
background geometry.
Therefore the Virasoro constraints can be rephrased as the invariance under the diffeomorphysms of
the coordinate s.

There are also another two  matrix model
realizations of effective actions involving the large N unitary
matrices. The first one looks as \cite{tierz}
\beq
Z_{CS}=\int dU e^{1/g_s Tr(logU)^2}
\eeq
and just this form makes the hidden quantum group structure
in CS theory quite manifest. The second model is useful for the
"crystallic" picture for Calabi- Yau and reads as \cite{okuda}
\beq
Z_{CS}= \int dU det \theta_{00}(U,q)
\eeq
where
\beq
\theta_{00}(e^{ix},q)= \sum_{m}q^{m^2/2}e^{imx}
\eeq
All matrix model realizations of the CS theory amount
to the same answer and can be equally used in the selfdual case.

It was discussed recently that one could expect two different matrix
realizations of the topological strings \cite{aganovic,rastelli,mal2004}.
One model can be attributed to the theory on the large number of branes
wrapped around the compact surface while the second Kontsevich type model involves the
finite number of the noncompact branes which deform the complex geometry.
Hence  we could look for the second Kontsevich realization of the effective
action in the selfdual field. To this aim it is convenient to use the
relation to the c=1 model at the selfdual radius mentioned above. The corresponding
Kontsevich models for this theory has been found in \cite{plesser,imbimbo}
and reads as
\beq
Z_{kon}(\mu,N)=\int dM e^{i\mu TrM - (i\mu +N)trlogM}
\eeq
The loop equation in this Kontsevich model or the corresponding
chiral ring relation provides
the conifold geometry where the type A model is defined on. Note that there
is the natural deformation of the c=1 partition function by the operators
corresponding to the tachyon modes. The emerging partition function
with "times" corresponding to the tachyonic     modes included can
be identified with the tau function of the Toda hierarchy. It would be
interesting to consider the corresponding deformations of the effective
action in the selfdual background. We plan to discuss this point elsewhere.

Let us emphasize that the matrix models we have discussed for the
selfdual background  have the same origin as  ones suggested
for the description of the Dirac operator spectrum in QCD
(see \cite{verba}, for a review).
Once again there are two different matrix model realizations.
The QCD is
essentially nonabelian theory and  the emergence
of the large N matrix model can be considered as a kind of averaging over
the instanton moduli space. This large N matrix model usually involves
the Gaussian measure. However there is the second Kontsevich type representation
in terms of unitary $SU(N_f)$ matrixes representing the Goldstone pion modes
\beq
Z_{QCD}=\int dU exp(\Sigma Tr(MU) +\nu logdetU)
\eeq
in the sector where $\nu$ is  the topological charge and $\Sigma$ is  the
chiral condensate.
The matrix resolvent has been identified with the resolvent of the
Dirac operator in this case too. It is known that the resolvent obeys
the algebraic equation defining the Riemann surface \cite{verba}.
It has the following structure
\beq
G^3(z) -2zG(z)+G(z)(z^2-\frac{1}{\Sigma^2}) - \frac{z}{\Sigma^2}=0
\eeq
and provides the geometry of the internal space
for the low energy QCD  similar to the discussion above.

\subsection{Spin 1/2 case}

Let us discuss the spinor case and
start with mentioning  the modifications which can be
expected from the very beginning.  First, note that
the loop representation for the effective action
implies that the spin factor corresponding to the WZW
action in $CP^3$ manifold has to be taken into account.
From the probe brane perspective the fundamental matter gets represented
by D7 branes  whose positions
fix the masses of the particles  and corresponding degrees of freedom
amounted from the strings connecting D3 and D7 branes.

The modifications of the type A and type B pictures
by the additional brane come
as follows. In type A picture the matter representing brane
yields the Wilson loop in Chern- Simons theory on $S^3$.
Such Wilson loops have been discussed in \cite{ooguri}
where the explicit answer for the corresponding
partition function has been derived. On the B model side
we have to insert the additional operators to the matrix
model realization. Recall the expression for the effective
action
\beq
L_{eff}=\int_{0}^{\infty}
\frac{ds}{s^3}e^{-sm^2} (esGcth(esG))^2
\eeq
and let us interpret it  in the proper way. First remind that the
factor $(esGcth(esG))^2$ is nothing but the product
of two massless propagators in $S^3$. Just these propagators
yield the Wilson loop in CS theory in type A picture.

In type B picture we deform the matrix model by the additional operators
and it is quite convenient to represent the effective action in the
following way
\beq
S_{eff}(m)=< \Psi_{2dYM}(s)|U(s,m)|\Psi_{2dYM}(s)>
\eeq
where we have exploited the relation between the particle
propagators  and the wave functionals in 2D YM theory
on the disc. In this case the Schwinger parameter plays
the role of the boundary holonomy which is the argument
of the wave functional. Hence we arrive at the following
picture; there are two discs with the single insertions
on each and SU(2) connection on each which are connected
by the cylinder. Each disc yields the corresponding wave function
and the integration over the Schwinger parameter corresponds
to the matrix element. Note that for the general spin J
the integrand has the additional spin  dependent terms
\beq
(Tr_{J}V(s))^2= (\chi_J(s))^2=(\frac{sin(2J+1)s}{sin s})^2
\eeq
which fixes the structure of the inserted operator.

The natural question concerns the modification of
the matrix model description in the spinor case. Hopefully
in the one-loop case the answer is very simple - there
is no need in the modification at all. The point is that
the selfdual case enjoys the residual supersymmety and the
one loop effective actions are related as follows
\beq
S_{1-loop,spin}= -2S_{1-loop,scal} + (\frac{ef}{8\pi})^2 log(m^2/\mu^2)
\eeq
The second term in the r.h.s. amounts from the fermionic zero modes
whose density in the selfdual fields is proportional to
$(ef)^2$. It is responsible for the chiral condensate
in the background field \cite{smilga}
\beq
m<\bar{\Psi}\Psi>= -\frac{(ef)^2}{8\pi^2}
\eeq
hence apart from zero modes the same matrix model can be used
for the spinor case.

Finally it is worth  questioning about the possible
worldsheet description of the spinor effective action.
To this  aim it is necessary to identify the operator
which has to be inserted on the topological string
worldsheet. It turns out that it can be recognized
in type A picture using
the example treated in \cite{berkooz} . It was shown that
the relevant deformation amounts to the multiplication of all string amplitudes
by the factors
\beq
exp(-c\int dx \sum_{kl} \Lambda_{k}(x,\bar{x})
\bar{\Lambda_{l}(x,\bar{x})} \frac{\partial}{\partial
\theta_k}\frac{\partial}{\partial \theta_l})
\eeq
where $\theta$
is the angular variable along the isometry in $S^3$.

Consider   two boundaries of the worlsheet localized in $S^3$ at
branes at points $z_i={\gamma_i,\bar{\gamma_i},\phi_i}$ where
the following string frame metric is implied here
\beq
ds^2=Q_5(d\phi^2 +e^{2\phi}d\gamma d\bar{\gamma})
\eeq
The
quasiclassical expression for the deforming operator reads as
\cite{kutasov}
\beq
\Lambda_i=-\frac{(\bar{\gamma_i}-\bar{x})e^{2\phi_i}}{1+
|\gamma_i-x|^2e^{2\phi_i}}
\eeq
If we assume that $\gamma_i=0$
then the emerging integral over the target
\beq
\int
d^{2}x|\frac{(\bar{x})e^{2\phi_1}}{1+ |x|^2e^{2\phi_1}} -
\frac{(\bar{x})e^{2\phi_2}}{1+ |x|^2e^{2\phi_2}}|^2
\eeq
amounts
precisely  to the factor $(-1 +(\phi_1 -\phi_2)ctg(\phi_1
-\phi_2))$ involved in the expression for the effective action in
the electric field.

Let as comment on the meaning of operator $\Lambda$. It is chosen
is such way that $\partial_{\bar{z}}\Lambda$ is a primary operator
of the worldsheet conformal algebra with dimension (0,1) as well
as a primary of space-time conformal algebra with dimension (1,0).
The insertion of the current $J(x)$ into the CFT correlator is
equivalent to an insertion of the vertex operator
\beq
K(x)\propto
\int dz k(z)\partial_{\bar{z}} \Lambda(z,x)
\eeq
in the string
worldsheet, where $k(z)$ is the worldsheet current \cite{kutasov}.
The worldsheet Lagrangian corresponding to this CFT
marginal deformation is nonlocal
\beq
\delta S\propto \int dx \int
dz_1 \int dz_2 k(z_1)k(z_2)\partial_{\bar{z_1}} \Lambda(z_1,x)
\partial_{\bar{z}} \Lambda(z_2,x)
\eeq
Let us emphasize that the worldsheet theory above involves
the nonlocal operators corresponding to the double trace
deformations so we meet here the touching surfaces once again.

\subsection{Magnetic selfdual field}

In the magnetic selfdual background the $S^3$ geometry
is substituted by the $AdS_3$ and the propagators
of the modes in $AdS_3$ are involved into the
effective action in the type A picture.  As we have
shown before the propagators in $AdS_3$ are related
to the integrated partition function of 2D SL(2,R)
Yang-Mills theory. Let us explain now that this
relation provides a proper type B picture.

To this aim let us remind that the topological SL(2,R)
Yang- Mills theory corresponds to the $AdS_2$ gravity
and argue that $AdS_2$ submanifold in $AdS_3$ is naturally
involved into the problem via the relation of 2D SL(2,R) YM theory
with perturbed topological gravity in two dimensions. It is known
\cite{FukKam85} that at $g^2 = 0$ SL(2,R) YM theory is equivalent to
the topological Jackiw-Teitelboim gravity with the action
\begin{equation}
S = \int d^2x \sqrt{det g_{\mu\nu}} \, \left( R(g) - \Lambda
\right) \eta \, ,
\end{equation}
where $\eta$ is the dilaton field and $\Lambda$ is the
cosmological constant. Solutions to the classical equations of
motion give rise to the $AdS_2$ gravity coupled to the dilaton.

The gravity degrees of freedom are zweibein $e^{a}(a=0,1)$ and
the spin connection $\omega$ which can be combined into the gauge
field \beq A=E^{a}P_{a} +\omega L \eeq where $P_a$ and $L$ are
the generators of the translations and Lorentz transformations
respectively. Due to the nonvanishing cosmological constant the Poincare
algebra is deformed to \beq [\Lambda,P_a]=\epsilon^{b}_{a}P_b
\qquad [P_a,P_b]= \epsilon_{ab}\Lambda L \eeq where $\Lambda$ is
the cosmological constant. This space can be identified with the
SL(2,R) algebra. The gauge curvature reads as \beq F=d\omega
+\Lambda e\wedge e \eeq and the condition of vanishing curvature
is nothing but the equation of motion in the Jackiw-Teitelboim model
\beq R=2\Lambda \eeq where R is the Ricci curvature scalar. Hence
the topological SL(2,R) 2D Yang-Mills theory
\beq
S_{top}=\int Tr\phi F
\eeq
describes the dilaton gravity in two dimensions.
The action enjoys the evident gauge invariance which is equivalent
to the general coordinate invariance in two dimensions.

The theory has no Hamiltonian and the Gauss law constraint
\beq
\partial \phi +[A_1,\phi]=0
\eeq
generating the gauge transformations  has to be imposed as
equation on the physical gauge invariant states in the Hilbert
space. It is clear that any functional of Wilson loop observable
obeys the quantum constraint which plays the role of the
Wheeler-De-Witt equation
\beq
 [\partial_{x}\frac{\delta}{\delta
A_{1}^{i}(x)} + \epsilon _{ij}^{l}A_{1}^{j}(x)
\frac{\delta}{\delta A_{1}^{l}(x)}]\Psi(A)=0
\eeq

The physical SL(2,R) Yang-Mills theory whose partition function is
involved into the effective action in the magnetic field
corresponds to the insertion of the operator \beq \delta
L=Tr\phi^2 \eeq in the action of worldsheet theory.  The origin
for this term to appear  has been explained in the
similar context in \cite{vafa}. Integration over the area
amounts to the peculiar wave functional in 2D gravity depending
on the boundary holonomy.

To match with the effective action picture we have to identify the
boundary holonomy in terms of the boundary paths. The comparison with
the propagator immediately gives
\beq \label{prop} trg=cosh(eHs)
\eeq
hence we have to check if this relation is consistent with
the worldvolume interpretation. Let us recall that the
holonomy of the SL(2,R) connection on a disc yields the length of
the boundary via the formula \cite{verlinde}
\beq
tr_{1/2}Pexp\oint_{C}A=2cosh\frac{l(C)}{2}
\eeq
Therefore we have
to check if the boundary length is proportional to $eEs$ indeed.
Qualitatively the proportionality to the Schwinger parameter is
correct since it defines the length of the boundary trajectory.

That is we have arrived at the proper gluing picture for
magnetic selfdual case as well. Indeed we have
qualitatively represented the effective
action of the spinor particle as gluing two $AdS_2$ manifolds
with fixed boundary length
perturbed by $Tr\phi^2$ operator each
\beq
S_{eff,magn}(m)=<\Psi_{SL(2,R)}(s)|U(s,m)|\Psi_{SL(2,R)}(s)>
\eeq
Note  that since $ADS_2$ has two boundaries there are some concerns
on this point which have to be clarified.

\section{One-loop low energy MHV amplitudes}

In this Section as a byproduct of the topological
string picture for the effective action in the selfdual field
found in this paper we shall develop the interesting
interpretation of
one-loop low energy  maximal helicity violating (MHV)
photon amplitudes. The idea to exploit the effective actions
to derive MHV amplitudes at one and two loops was used
in \cite{shubert2,dunne2}. The key point is to consider
the limit when all photon momenta are small
compared to the mass of the particle in the loop. In this
limit the effective action serves as the generating function
for the amplitude with the arbitrary number of the external photon
legs. To derive the amplitude from the effective action one
introduces the momenta and polarization for each external leg
\beq
F^{\mu\nu}_i=k^{\mu}_i\epsilon^{\nu}_i- k^{\nu}_i\epsilon^{\mu}_i
\eeq
It is convenient to define
\beq
F_t=\sum_{i=1}^{N}F_i
\eeq
and expand the one loop effective action in powers of $F_t$.
To get the amplitude with N external photons the effective
action has to be expanded to the N-th power in the external
field $F_t$ and only terms which involve each $F_i$ linearly
are kept. To obtain the MHV amplitudes when polarizations
of all photons are "+" or "-" it is necessary to consider
the effective action in the selfdual background since selfdual fields
have fixed chirality \cite{duff}. The expansion
of the effective action amounts to the following answer for the
N photon low energy amplitudes at one loop in QED
\beq
\Gamma^1(k_i,\epsilon_i)= -\frac{2(2e)^N}{(4\pi)^2m^{2N-4}}c^1(N/2)\chi_N
\eeq
where
\beq
c^1(n)=-\frac{B_{2n}}{2n(2n-2)}
\eeq
$B_n$ are Bernoulli numbers and $\chi_N$ is kinematical factor
which can be presented in the spinor helicity notations
\beq
\chi_{N}=\frac{(N/2)!}{2^{N/2}}([12]^2[34]^2\dots [(N-1)N]^2 +perm)
\eeq
Similar expression can be found for scalar electrodynamics and
the two-loop generalization has been derived \cite{dunne2}.

Turn now to the topological string interpretation of such MHV
amplitudes. To this aim let us use the relation to large N CS theory
discussed above and expand $log Z_{CS}$ in inverse powers of level k.
The N-th term $\cal{F}_N$ in the expansion corresponds to the
N-th term in the expansion of the effective action in the selfdual
background that is it corresponds to the amplitude with N external
legs. On the other hand CS theory can be mapped into the
topological string amplitudes and the l-th term in the expansion
of the CS partition function corresponds to the genus zero
topological string amplitude with l holes corresponding
to the insertion of the vertex operators.

The situation has many common features with the recent approach
to the calculation of the MHV amplitudes in N=4 SYM theory \cite{wittentw}.
In nonabelian case the all "+" MHV one-loop massless  MHV amplitudes were
calculated both in YM theory \cite{bern} and massless QED \cite{mahlon}.
These amplitudes are  related to the amplitudes
of topological type B strings which are localized on the holomorphic
genus zero curves in twistor target space \cite{one-loop}.
The Minkowski coordinate parametrizes the moduli of the curve in
the twistor space. Moreover as amplitude is localized on the
complex line in the twistor space it corresponds to the
point-like vertex in the Minkowski space.

In our abelian example  the answer can be reformulated
in terms of the topological type B  string amplitudes
in Calabi-Yau manifold once again. In our case the mirror of $T^{*}S^3$
plays the role of the "twistor" manifold and the mass parametrizes the
geometry. Hence we have here some analogue of the "twistor" manifold
for nonsupersymmetric case involving the mass parameter.
In this low-energy limit we have the transparent localization of the amplitudes to the point
in the Minkowski space. Indeed
the effective action in the selfdual field
serves as the generating function for all "+"  low-energy MHV
amplitudes. On the other hand in the low energy limit of
the large mass to get the amplitudes we just expand the effective
action in terms of the set of local operators multiplied
by $m^{-2N}$. In the first quantized formulation it can be thought of
as the expansion of the Wilson loops in terms of the local operators.
Therefore the low-energy amplitudes are presumably localized on the complex
lines in the "twistor" manifold.

The advantage
of the massive case here is that the topological string
picture can be extended to the higher loops immediately. We believe that the abelian
low energy one-loop MHV amplitudes could serve as a good toy model for the topological
string interpretation
of the nonabelian MHV amplitudes in nonsupersymmetric case.
Note also that we have type A picture for the amplitude as well
described by large N Chern-Simons theory or closed topological theory
upon the large N geometrical transition. In particular the mass parameter
effectively measures the Kahler class of $P^1$ in the resolved conifold.
Since we have developed the matrix model description for type B topological strings
corresponding to the effective actions it can be used
for the calculation of the one-loop MHV diagrams. The k-th terms of the expansion
of the matrix model partition function correspond to the one-loop
abelian MHV amplitudes with k external legs.

\section{Imaginary part}

In this Section we shall briefly comment on the relation of the $S^3$
geometry with the Schwinger pair production. In the electric field
the effective action develops the imaginary part which is
responsible for the probability of the pair creation. The
probability  looks as \cite{schwinger}

\beq w=(eE)^2 \; \frac{2s+1}{8\pi^2}\sum_{n=1}^{\infty}
\frac{(-1)^{(2s+1)(n+1)}}{n^2} \exp\left(-\frac{n\pi
M^2}{eE}\right) \eeq where $s=0$ or $1/2$ is the particle spin,
$e$ -- charge and $E$ -- constant electric field. The
corresponding stringy generalization has been found in \cite{bp}
and the relevant brane geometry in \cite{gss}.

Let us recall the geometrical meaning of the Schwinger parameter
in the calculation of the imaginary part and discuss the leading
exponential factors first. The imaginary part in the path integral
is saturated by the classical trajectories in the Euclidean space.
Since upon the Wick rotation the electric field gets transformed
into the magnetic one the trajectories are circles. To get the radii of
the circles one could first integrate over the proper time in the
path integral which amounts to  the
Schwinger parameter $s=L/m$, where L is the length of the contour.
The radii can be obtained by the  minimization of the effective
action for the negative radial mode (see, for
instance, \cite{affleck}) \beq S_{eff}= mL-eEA \eeq where $L$ and
$A$ are the perimeter and the area of the closed particle
Euclidean trajectory. Then the radius can be found and the
probability up to the exponential accuracy reads as \beq w \sim
\exp (-S_{eff}^{min})
\eeq
Thus the Schwinger parameter takes the
quantized values at the leading approximation.

Let us examine now the probability of this process from the $S^3$
geometry viewpoint. It is clear that the imaginary part in the
effective action amounts from poles in the Schwinger parameter
integration located at \beq eEs_k=2\pi k \eeq  Remind that
we have related the  Schwinger parameter with the group coordinates
via the relation $trg=cos esH$. Remarkably enough these poles
correspond precisely to the possible quantized positions of $S_2$
branes in $S_3$ \cite{bachas}. These branes are stable due to the
fluxes on their worldvolumes. Hence we could suggest that these
SU(2) branes are involved into the tunneling process.
The possible mechanism  mentioned in \cite{bachas} deals with the decay of D0 branes located at
the group unit to the extended brane located at $s_k$.

After the Euclidean rotation to $AdS_3$
the tunneling process can be embedded into the string picture as follows.
The Euclidean circle above is just one of the boundaries of the cylinder
extended along the radial $AdS_3$ coordinate while the length
of the cylinder is proportional to the particle mass hence
the area of the cylinder is just the first term in the effective
action \cite{gss}. On the other hand the same geometry can be considered
as the propagation of the rigid closed string in the different
channel. Then the Euclidean particle trajectory is just the
fixed time closed string. Therefore in this channel the
poles in the $AdS_3$ propagator can be identified with
the closed string states.

It could be questioned how the arguments above
can be extended to the strong coupling region and match the
calculation of the vacuum expectation values of the Wilson loops
in the strong coupling limit \cite{wilson}. In that case the
relevant surface has the disc topology with Wilson loop boundary. To
get  transition from the cylinder to disc topology let us
remind the interpretation of the critical electric field found in
\cite{gss}. Namely it was shown that when the finite string tension is
taken into account the minimal surface deviates from the cylinder
form and at the critical field
\beq eE=\frac{1}{\alpha '} \eeq
the
surface shrinks at one point transforming into two disconnected
surfaces. Hence in the closed string channel the strong coupling
calculation of the Wilson loops corresponds to the overcritical
electric field.

\section{On the geometry of paths}

Since the effective actions can be represented via the sum over
the closed paths weighted with the proper factors it is natural to
ask which paths really dominate the sum. Here we would like to
present a couple of arguments which imply that the paths with the
cusps are relevant. The first argument involves the observation
made above that the double-trace deformations of the worldsheet
action turns out to be important. Moreover
the matrix model describing the effective action in the selfdual
case involves the double trace terms in the potential as well.
It is wellknown that the
double-trace deformations correspond to the touching of the string
worldsheets.

The second argument involves the restrictions on the invariant
lengths of the particle trajectories. The simplest example which
one could have in mind involves the temperature. Namely,
consider the classical particle trajectories in the Euclidean
space-time in the electric field. If there is no temperature then
the classical trajectories are just the circles. When the
temperature is switched on  the periodicity condition in the
Euclidean time should be imposed. Hence there are two different
situations with the radius of the circle corresponding to the
classical motion is larger or smaller then $T/2$. In the former
case we have specific situation when the circle becomes two arcs
instead  and the cusps emerge immediately.

In our case there is no temperature and therefore periodicity in
the Euclidian time direction. However we have to integrate over
the invariant lengths of trajectories and if the total length of
the trajectory is fixed by the value of the Schwinger parameter
then it imposes the restriction on the contour. Hence we are
tempting to speculate that in the Euclidean geometry we have two
arcs once again and the angle at the cusp is fixed by the length
on the trajectory.

The last argument concerns  the two loop calculation of the effective
action which generically  provides the information
about one loop anomalous dimensions of
operators. It is convenient to
compare the  structure of the two-loop answer with the
cusp anomalous dimensions representing the renormalization of the
Wilson loop with cusps. It was calculated long time ago
\cite{Polyakov1} and reads as
\beq
\Gamma_{cusp}(\alpha,\theta)=\alpha_s N(\theta coth\theta -1) +
O(\alpha^2)
\eeq
It corresponds to the renormalization of the
Wilson loop with cusp angle $\theta$ and serves
as the generating function for the anomalous
dimensions of the operators with the large Lorentzian spin.

Let us look more carefully
at the two loop answer and focus for a moment on the theory with N=2 SUSY.
Effective action in the SUSY case has the following structure in two
loops \cite{kuzenko}
\beq
S_{eff,2loop}=g^4 \int  ds f(s)(es\Psi
cth(es\Psi) )
\eeq
where f(s) is the some polynomial and $\Psi$ is
N=2 superfield. One immediately recognizes the cusp anomalous
dimension which  implies that the renormalization of the Wilson
loop with the cusp indeed is relevant. On the other hand
the two-loop effective action in the selfdual background
can be expressed through one-loop answer \cite{dunne2}
providing one more argument along this line.

We have argued above that it is natural to conjecture that the
contours with cusps are relevant for the representation of the
effective action in the first quantized picture. On the other hand
we have found  some counterpart of the cusps in the dual stringy
picture via the double trace operators. Of course the arguments
above are  quite tentative hence the additional clarification
of this issue is needed.

\section{Discussion}

In this paper we have analyzed to what extend the one-loop
effective actions in the nonsupersymmetric theory
feel the higher dimensional background
geometry. This is  among the first steps which would provide
the stringy picture for the perturbative regime in the
nonsupersymmetric gauge theory. It turns out that the
effective actions  amount
to  the relatively precise background geometry which depends on
the choice of the abelian background field. Purely electric
field feels the SU(2) part of background,  purely magnetic case
involves the $AdS_3$ geometry while the selfdual background
fixes the full Calabi-Yau threefold geometry. The key point
of  such identification is the interpretation
of the Schwinger parameter emerging in the loop calculation
as the coordinate in the background geometry. Then the
integrands in the loop integrals are nothing but the
propagators of the different massive or massless modes
in this geometry.

The most transparent picture emerges for the selfdual
background when the full three dimensional complex manifold
has been determined. It turns out that the link to the
large N CS theory fixes it to be $T^{*}S^3$ or its mirror.
The effective action in this case can be related to the
topological string amplitude in type A or B models.
Moreover it turns out to be possible to develop the matrix
model description of the selfdual case which is some sence
can be considered as the counterpart of Dijkgraaf-Vafa
picture for the nonsupersymmetric case. The  important point
is that at least one  matrix model involves a set of double trace
operators which differ the model from the N=1 SUSY case.

As a byproduct we have found the interesting interpretation of the
abelian MHV  loop amplitudes in terms of the topological type
B strings. The target space for the topological strings has been
defined and the amplitudes can be calculated or from the perturbative
expansion in CS theory on the A model side either in the matrix model
on the B model side.
However in spite of the evident similarity with the twistor
type picture for the nonabelian massless case the additional work is needed to
clarify the possible interpretation of the mirror to $T^{*}S^3$ as a kind
of the twistor manifold. Note that the  phenomena of the
generation  of the local vertexes for the MHV amplitudes
is transparent in the low energy limit.

Our analysis strongly suggests that the closed loops
with self intersections which are holographycally dual
to the multitrace operators in the bulk are important in
the path integral. Moreover the geometry is encoded in the
loop equations for the Wilson loops and Ward identities
for the integral over Schwinger parameter. We also discuss
some features of the nonperturbative pair production
in terms of background geometry.
Let us emphasize that the nontrivial background geometry discovered
behind the one-loop effective actions can not be globally seen at any finite order in the
external field. However it is important that the gauge coupling
can be made arbitrary small hence the emergence of the nontrivial
background is a weak coupling phenomena.

In our abelian nonsupersymmetric case
the external field plays the  role
similar to the  composite field $S=TrW^2$ in N=1 theory. For the
nonabelian nonsupersymmetric case the role of S is played by
the composite colorless field $\sigma=TrF^2$ so the derivation
of the Veneziano-Yankielowicz effective Lagrangian \cite{veneziano} via
matrix model representation of the resolved conifold can be
parallelled with the effective potential in nonsupersymmetric case
found in \cite{shifman}. Note also that $\beta$ function of the
theory can be derived from the effective action in the strong field limit
hence the difference between the signs in the abelian and nonabelian
cases acquires the geometrical interpretation. It would be interesting
to combine our geometric interpretation of one-loop answer with
the explicit instanton counting in N=2 case \cite{nekrasov}.

It is clear that there are a lot of questions to be answered
within our approach. Some of them have been already mentioned in
the body of the paper. The most immediate one concerns the
nonabelian generalization of the geometrical picture discussed
in this work. We shall discuss the associated one-loop
nonabelian phenomena including Nilesen-Olesen instability
and low-energy effective Lagrangian in a furthercoming publication.
We shall also discuss the relation of the higher loop corrections
to the effective actions to the  topological string
interpretation of the low energy abelian MHV
amplitudes at higher loops.

The work was partially supported by grants RFBR-040100646 (A.G) and
RFBR-040216538(V.L). (A.G.) thanks
Seoul National University  where the part of the work has been
done for the  hospitality. He also thanks organizers of the
program "QCD and String Theory"  at Kavli Institute for Theoretical Physics  at
UCSB supported by grant NPP-199-07949
and FTPI at University of Minnesota where
it has been completed for the hospitality and support.

\section{Appendix A}

We shall argue now that the massless  propagator in the $AdS_3$ can be
formulated in the second quantized 2D worldsheet picture as two
point function (see the discussion in \cite{gorsky}).
In this approach instead of the inserting of the  double trace operator
on the worldsheet we choose the Rindler vacuum state.
To this aim
consider the two dimensional worldsheet field theory  with the
equation of motion
\beq (\partial_{t}^2 -\partial_{x}^2)\phi +m^2\phi=0 \eeq whose
solution has the following mode expansion
\beq
\phi(x,t)=\int \frac{d\beta}{2\pi}
(a^{*}(\beta)e^{-im(xsinh\beta -tcosh\beta} +
a(\beta)e^{im(xsinh\beta -tcosh\beta})
\eeq

It is convenient to introduce Rindler coordinates
\bea
x=rcosh\theta,\qquad t=rsinh\theta
\nonumber\\
-\infty < \theta < +\infty \qquad 0 < r <+\infty
\eea
in the
space-time region $x>|t|>0$. Let us perform the following Laplace
transform with respect to the radial coordinate
\beq
\lambda_{\theta}(\alpha)=\int dr
e^{imrsinh\alpha}(-\frac{1}{r}\partial_{\theta} +
imcosh\alpha)\phi(r,\theta)
\eeq
Then the commutation relation for
the Laplace transformed field reads as
\beq
[\lambda(\alpha_1),
\lambda(\alpha_2)]=i\hbar tanh(\alpha_1 -\alpha_2)/2
\eeq
and the
Hilbert space is spanned by vectors $a(\beta_n)\dots
a(\beta_1)|vac>$ where the vacuum state is defined as
\beq
a(\beta)|vac>=0 \qquad <vac|a^{+}(\beta)=0
\eeq
One can introduce the
two point function
\beq
F(\alpha_1-\alpha_2)=
<vac|\lambda(\alpha_1)  \lambda(\alpha_2)|vac>
\eeq
and it appears
that the explicit calculation amounts to the following answer
\cite{luk}
\beq
F(\alpha -i\pi)= - \frac{1}{\pi}\alpha /2 coth(\alpha /2) +
singular \quad terms
\eeq
The singular terms cancel in the
difference $ F(\alpha -i\pi) - F( 0)$ which coincides with the
propagator of the massless mode in $AdS_3$.

\section{Appendix B}
Let us consider the Lagrangian of the scalar field in the constant
magnetic field
\beq
L=-(\partial_\mu+ieA_\mu)\phi^*(\partial_\mu-ieA_\mu)\phi
\eeq
and take $A_1=A_2=A_4=0, \, A_2=Hx_1$. Then the eigenvalue
equation becomes \beq
-E^2\phi_E =
(\triangle-e^2H^2x_1^2-2ieHx_1\frac{\partial}{\partial x_2})\phi_E
\eeq
yielding the  solution to the eigenvalue problem
\beq
E^2=2eH(n+\frac{1}{2}+k_3^2)
 \eeq
The same calculation can be done for arbitrary spin and the answer
will be the following
\beq E^2=2eH(n+\frac{1}{2}+S_3)
\eeq
hence the sum
of all zero-point energies becomes
\beq F=c\int dk_3 \sum_{n,S_3}
\sqrt{2eH(n+\frac{1}{2}+S_3)+k_3^2}
\eeq
In general this sum is
divergent,  however performing the $\epsilon$ regularization we obtain \beq
F=\mu^{-2\epsilon}(const)\int^\infty_0 d\tau
\tau^{-2-\epsilon}\sum_{n,S_3}e^{-i\tau 2eH(n+1/2+S_3)} \eeq
amounting
after the  rotation of the contour to
\beq
F=\mu^{-2\epsilon}(const)\int_0^\infty \frac{ds}{s^{2+\epsilon}}
\frac{sinh(2S+1)eHs}{sinh^2(eHs)}
\eeq

\end{document}